\documentclass[twocolumn,superscriptaddress,showpacs,showkeys,aps,prl]{revtex4}

\usepackage[final]{graphicx}
\usepackage{t1enc}
\usepackage{bm}

\begin{document}
\bibliographystyle{apsrev}

\title{Emergent self-organized complex network topology out of stability constraints}

\author{Juan I. Perotti}
\affiliation{Instituto de F\'\i sica de
la Facultad de Matemática, Astronomía y F\'\i sica
(IFFAMAF-CONICET), Universidad Nacional de C\'ordoba,
\\ Ciudad Universitaria, 5000 C\'ordoba, Argentina}

\author{Orlando V. Billoni}
\affiliation{Instituto de F\'\i
sica de la Facultad de Matemática, Astronomía y F\'\i sica
(IFFAMAF-CONICET), Universidad Nacional de C\'ordoba,
\\ Ciudad Universitaria, 5000 C\'ordoba, Argentina}

\author{Francisco A. Tamarit}
\affiliation{Instituto de F\'\i
sica de la Facultad de Matemática, Astronomía y F\'\i sica
(IFFAMAF-CONICET), Universidad Nacional de C\'ordoba,
\\ Ciudad Universitaria, 5000 C\'ordoba, Argentina}

\author{Dante R. Chialvo}
\affiliation{Department of
Physiology, Northwestern University, Chicago, IL 60611, USA}
\author{Sergio A. Cannas}
\email{cannas@famaf.unc.edu.ar}  \affiliation{Instituto de F\'\i
sica de la Facultad de Matemática, Astronomía y F\'\i sica
(IFFAMAF-CONICET), Universidad Nacional de C\'ordoba,
\\ Ciudad Universitaria, 5000 C\'ordoba, Argentina}
\begin{abstract}
Although most networks in nature exhibit complex topologies the
origins of such complexity remain unclear. We propose a general evolutionary mechanism based on global stability. This mechanism is incorporated into a model
of a growing network of interacting agents in which each new
agent's membership in the network is determined by the agent's
effect on the network's global stability. It is shown that out of
this stability constraint,  complex topological properties emerge in a self
organized manner, offering an explanation for their observed
 ubiquity  in biological networks.
\end{abstract}


\maketitle

Complex networks of interacting agents are ubiquitous, in a wide
range of scales, from the microscopic level of genetic, metabolic
and proteins networks to the macroscopic human level of the
Internet \cite{AlBa2002,SoFeMoVa2003}. All of them exhibit high
clustering and relatively short path length compared with random
networks. In addition, they frequently show a nonhomogeneous structure, characterized by a degree distribution (the probability of a node to be connected to $k$ other ones) with a broad tail
 $P(k) \sim k^{-\gamma}$ for large values of $k$, with
exponents $\gamma<3$
\cite{AlBa2002,SoFeMoVa2003, Sporns2004}. Several mechanisms have been
proposed to give rise to this kind of topologies\cite{AlBa2002,Baiesi2003}.
These mechanisms have successfully explained the origin of complexity
in some networks, but it is recognized that another, equally
large, number of cases can not be accounted for either class of
models. In particular, growing biological networks involve the
coupling of at least two dynamical processes. The first one
concerns the addition of new nodes, attached either during a slow
evolutionary (i.e., species lifetime) or a relatively faster
developmental (i.e., organism life time) process. A second one is
the node dynamics which affects and in turn is affected by the
growing processes. It is reasonable to expect that the
network topologies we finally witness could have emerged
out of these coupled processes. This
Letter is dedicated to discuss a simple model of this problem,
showing that complex networks do emerge under general realistic
constraints. It
needs to be noted from the outset that the aim of this letter
is not to describe an arbitrary algorithm, but to identify a
dynamical process able to be implemented by natural systems.

Before introducing the model, and to fix ideas, let us dwell on
some concrete general examples. First consider a food web, which
is constructed through community assembly rules, strongly
influenced by the underlying dynamics of species and specific
interactions among them\cite{WeKe1999,Pi1991}. Another example could be
neuronal networks, where the addition of hundreds of thousands of new
neurons is followed by a dynamical process in which neuronal
dynamics and connectivity are interrelated in a way not fully
understood. Yet a third example at another scale, could be
imagined in the context of social networks, in which novice
members can be accepted or rejected based on their individual
contribution to a global interest, fitness, performance or profit.
In the three examples it is relatively easy to visualize the two
processes mentioned above. The consequence of adding a new member
with a given connectivity affecting a global in/stability, is
represented in these examples by the aboundance/lack of food, the
neuronal welfare/death or the profits' up/down, respectively. Notice that each new member may not only result in its own
addition/rejection to the system, but it can also promote avalanches
of extinctions amongst existing members, an effect we found that
strongly influences the network's topology.

Let us consider a system of $n$ interactive agents, whose
dynamics is given by a set of differential equations $d \vec{x}/dt= \vec{F}(\vec{x})$, where $\vec{x}$ is an $n$-component vector describing
the relevant state variables of each agent and $\vec{F}$ is an
arbitrary non-linear function.  One could imagine that $\vec{x}$ in
different systems  may represent concentrations of some hormones,
or the average density populations in a food web, or the
concentration of a chemicals in a biochemical network, or the
activity of  genes in a gene regulation net, etc. We  assume that
a given agent $i$ interacts only with a limited set of $k_i < n$
other agents; thus $F_i$ depends only on the variables belonging to that
set. This defines the interaction network, as was done
previously\cite{Ma1972}.

We will assume that there are two time scales in the dynamics. On
the long time scale (much larger than the observation time) the
system is subjected to an external flux (migration, mutation,
etc.) of new agents that interact with some of the previous ones
and can be incorporated into the system or not, so $n$ (and the
whole set of differential equations) can change. On short time
scales we assume that $n$ is constant and the dynamics already
led the system to a particular stable stationary state
$\vec{x}^*$ defined by $\vec{F}(\vec{x}^*)=0$ \cite{NoteLett0}. The stability of
that solution is determined by the eigenvalue with maximum real
part of the Jacobian matrix $a_{i,j}\equiv \left(\frac{\partial
F_i}{\partial x_j}\right)_{x^*}$. Therefore a new agent will be
incorporated to the network if its inclusion result in a new stable
fixed point, that is, if the values of the interaction matrix
$a_{i,j}$ are such that the eigenvalue with maximum real part
$\lambda$ of the enlarged Jacobian matrix is negative ($\lambda<0$). Assuming
that isolated agents will reach  stable states by themselves after
certain characteristic relaxation time, the diagonal elements of
the matrix ${a_{i,i}}$ are negative and given unity value to
further simplify the treatment\cite{Ma1972}. The interaction values, (i.e.,
the non-diagonal matrix elements ${a_{i,j}}$) will take random
values (both positive and negative) taken from some statistical
distribution.  In this way we have an unbounded ensemble of systems\cite{Ma1972} characterized by a ``growing through stability'' history. Randomness  would be self-generated through the addition of new agents processes. Each specific set of matrix elements after addition defines a particular dynamical system and the subsequent analysis for time scales between successive migrations is purely deterministic.

These ideas are implemented in a numerical model as follows: At
every step the network can either grow or shrink. In each step an
attempt is made to add a new node to the existing network,
starting from a single agent ($n=1$).  Based on the stability
criteria discussed, the attempt can be successful or not. If
successful, the agent is accepted, so the existing $n \times n$
matrix grows its size by one column and one row. Otherwise the
novate agent will have a probability to be deleted together with
some other nodes as further explained below.
 More specifically, suppose that we have an already created network
with $n$ nodes, such that the $n \times n$ associated interaction
matrix ${a_{i,j}}$ is stable. Then, for the attachment of the
$n+1_{th}$ node we first choose its degree $k_{n+1}$ randomly between $1$
and $n$ with equal  probability. Then the new agent interaction
with the existing network member $i$ is chosen such that non-diagonal
matrix elements $(a_{i,n+1},a_{n+1,i})$ ($i=1,\ldots,n$) are zero
with probability $1-k_{n+1}/n$ and different from zero with probability
$k_{n+1}/n$; to each non--zero matrix element we assign a different real random value uniformly
distributed in $[-b,b]$. $b$ determines the interaction range
variability and it is one of the two parameters of the model.
Then, we calculate numerically $\lambda$ for the resulting $(n+1)
\times (n+1)$ matrix. If $\lambda<0$   the
new node is accepted. If $\lambda>0$  
it means that the introduction of the new node destabilized the
entire system and we will impose that, either the new agent is
eliminated or it remains but produces the extinction of a certain
number of previous existing agents.
In order to further simplify the numerical treatment, we will
allow up to $q \leq k_{n+1}$ extinctions, taken from the set of
$k_{n+1}$ nodes connected to the new one\cite{NoteLett1}; $q$ is
the other parameter of the model. To choose which nodes are to be
eliminated, we first select one with equal probability in the set of
$k_{n+1}$ and remove it. If the resulting $n \times n$ matrix is
stable, we start a new trial; otherwise, another node among the
remaining $k_{n+1}-1$ is chosen and removed, repeating the
previous procedure. If after $q$ removals the matrix remains
unstable, the new node is removed, we return to the original
$n\times n$ matrix and start a new trial\cite{numerics}.

First we calculated the average connectivity $C(n)$, defined as
the fraction of non-diagonal matrix elements different from zero,
averaged over different runs.  We found that $C(n)\sim
n^{-(1+\epsilon)}$ (see Supplementary information) for large values of
$n$, where the exponent $\epsilon$  depends on $b$ and $q$, taking
values $0<\epsilon<1$. Such behavior is characteristic of food
webs\cite{SoGo2000} and it has been interpreted in terms of
self-organized criticality concepts\cite{SoAlMc2000}; the present
results suggest that this is a general behavior in
stability-driven self organized systems.

\begin{figure}
\begin{center}
\includegraphics[scale=0.27]{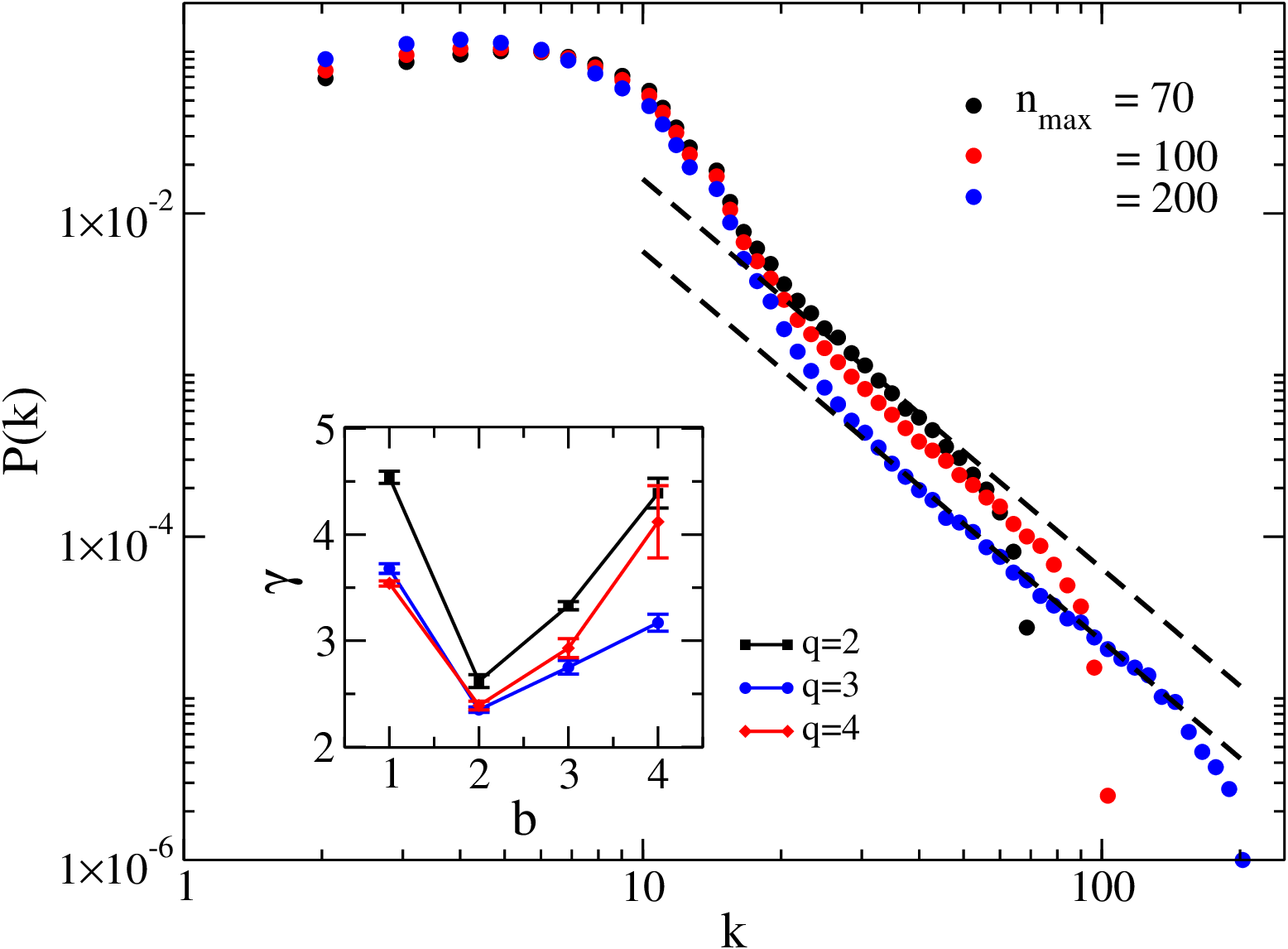}
\caption{\label{fig1} (Color on-line)  Degree distribution $P(k)$
for  $b=2$, $q=3$ and different values of $n_{max}$; the dashed
lines correspond to a power law $P(k) \sim k^{-\gamma}$ with
$\gamma \approx 2.4$.  The inset shows $\gamma$ as a
function of $b$ for different values of $q$.}
\end{center}
\end{figure}

Next we calculated the degree distribution $P(k)$ of the network
with $n=n_{max}$ for different values of $b$ and $q$. The typical
behavior of $P(k)$ is illustrated in Fig.\ref{fig1}  for $b=2$,
$q=3$ and different values of $n_{max}$. We see the
emergence of a fat tail $P(k) \sim k^{-\gamma}$ for large
values of $n$, with an exponent $\gamma$,
independently of the network size (this figure also shows that the
drop in the tail of the distribution is a finite size effect). Notice that this relatively small range of the broad tail is what more often is seen in real networks. The qualitative behavior of
 $P(k)$ for other values of $b$ and $q$ is the same (see Supplementary information). The inset of Fig.\ref{fig1} shows the value of the exponent
$\gamma$ as a function of $b$ for different values of $q$. We see
that $\gamma$ presents a minimum around $b=2$ for all values of
$q$; as $q$ increases the exponent  decreases and
for large enough values of $q$ we obtain a non-trivial value of
$\gamma<3$ for a broad range of values of $b$.

To exclude the possibility that the observed network topology is
trivially associated with a hidden preferential attachment (PA)
process, we computed the attachment probability $\Pi(k)$, defined
as the probability that a new node connects with an already
existing node with degree $k$. Assuming that the average degree
$\left< k_i\right> \ll n$, the attachment probability can be
expressed as $\Pi(k)=\sum_i^{n_k} \Pi_i$, where $\Pi_i$ is the
probability that the new node connects to the already existing
node $i$, $n_k \approx n\, P(k)$ is the number of nodes with
degree $k$ and the sum runs over all sites $i$ with degree
$k_i=k$. If stability selection would favor some kind of
PA mechanism, (i.e., if new nodes are
attached with larger probability to nodes highly connected)  we
should expect
$\Pi_i = \frac{k_i}{\sum_{j=1}^n k_j} \approx \frac{k_i}{ n\,(n-1)
C(n)}$

\noindent and therefore

\begin{equation}
\Pi(k) \approx \frac{P(k)\, k}{ \,(n-1) C(n)}\label{Pik}.
\end{equation}

\begin{figure}
\begin{center}
\includegraphics[scale=0.27]{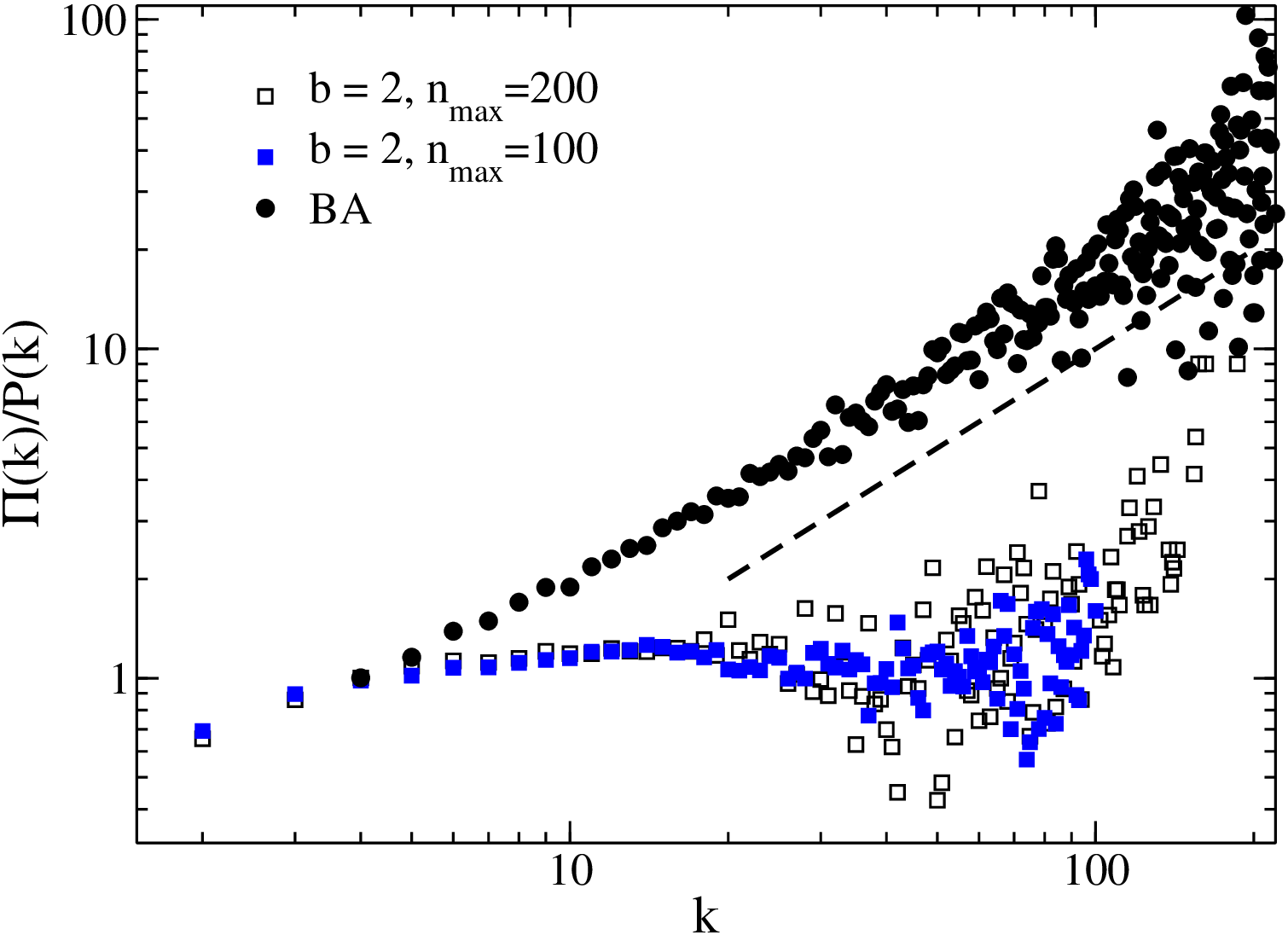}
\caption{\label{fig2} (Color on-line) Relative attachment
probability $\Pi(k)/P(k)$ for  different values of $n_{max}$,
compared with the corresponding results for a BA network of the
same size. Dashed line corresponds to a linear behavior
$\Pi(k)/P(k)\sim k$.}
\end{center}
\end{figure}

In Fig.~\ref{fig2} the relative attachment probability
$\Pi(k)/P(k)$ in the present model for a fixed network size $n$
and different values of $b$  is compared with the corresponding
results for a network of the same size obtained with the
Barab\'asi-Albert (BA) \cite{AlBa2002} algorithm with connectivity
$C(n)$. This quantity shows the expected
behavior $\Pi(k)/P(k) \sim k$  for large values of $k$,
consistently with Eq.(\ref{Pik}). In  the present model $\Pi(k)/P(k)$ remains almost constant for a wide
range of values of $k$ (including a range of values for which the
power law behavior of $P(k)$ has already established), but
displays  an increasing trend consistent with Eq.(\ref{Pik}) for
large values of $k$. In other words, in the present model at
variance with the BA model, as the network grows, the assembly
mechanisms selected by stability shows a crossover between two
regimes: one dominated by PA and the other
not.

Considering that biological systems are probably never in a
completely stable situation, we relaxed the condition of stability
$\lambda <0$ and look at networks growing by allowing $\lambda$ to
take small positive values so that the characteristic time to
leave an unstable fixed point $\tau=\lambda^{-1}\gg 1$. By
accepting nodes as long as $\lambda < \Delta$ the calculation of
$P(k)$ for different values of $\Delta$ (positive and negative)
showed similar qualitative behavior, with small variations of the
$\gamma$ exponent (See Supplementary Information).

\begin{figure}
\begin{center}
\includegraphics[scale=0.44]{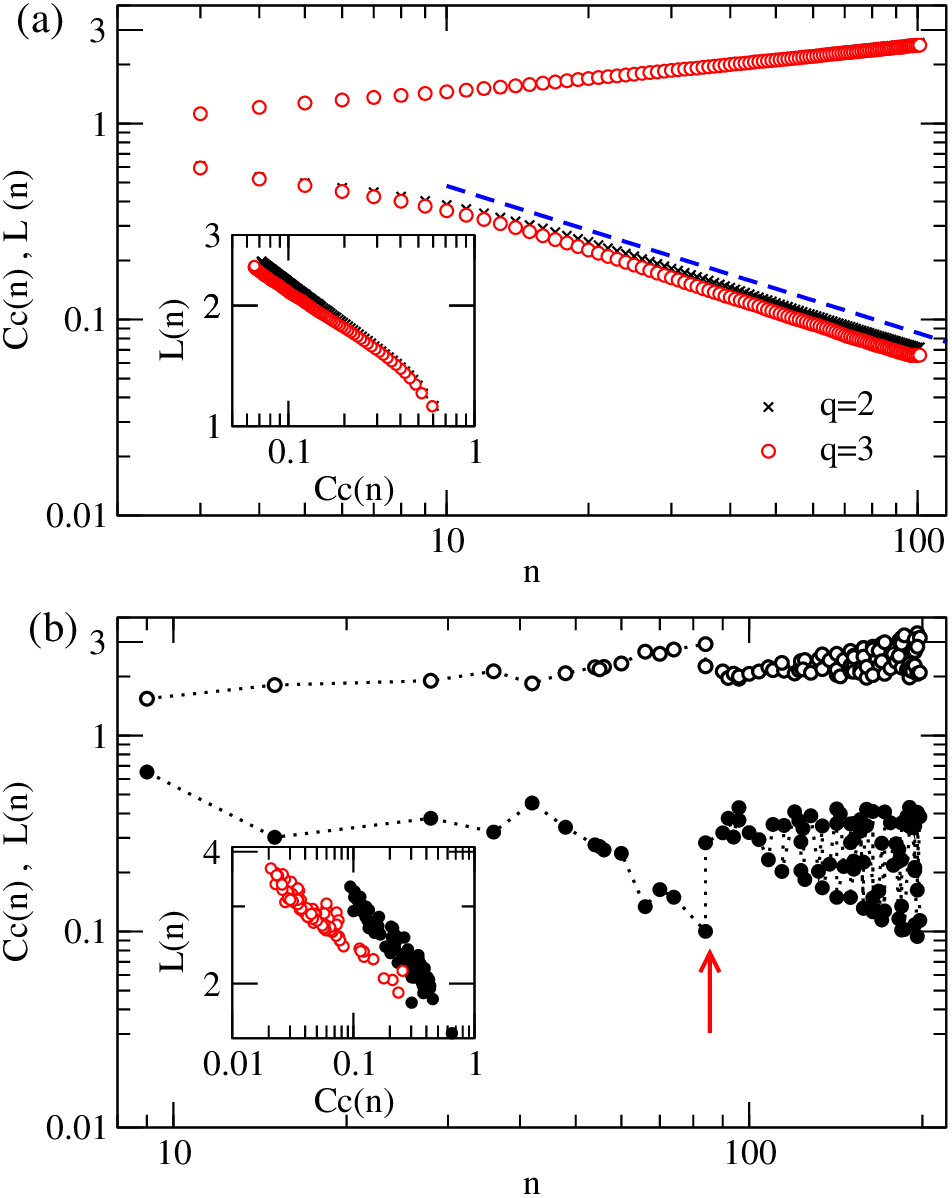}
\caption{\label{fig3} (Color on-line) (a) Networks' average
clustering coefficient $Cc(n)$ and $L(n)$ for $b=2$ and different
values of $q$ as  functions of network size. The dashed line is a
guide to the eye corresponding to $Cc(n) \sim n^{-0.75}$. Inset
shows the same data plotted against each other. (b) $Cc(n)$ and
$L(n)$ as  functions of network size computed from a single
network realization. Data are samples taken every fifty trials,
regardless of the resulting stability. Notice how fluctuations
increase as the network grows. Inset shows the same data plotted
against each other (full circles), in addition to the data computed from a random
network with equal size and density of connections (open circles).}
\end{center}
\end{figure}

Next we calculated the average path length $L$ between two nodes
and the average cluster coefficient $Cc$ for the networks obtained
by the present algorithm as a function of the network size $n$.
$L$ is defined as the minimum number of links needed to connect
any pair of nodes in the network and $Cc$ is defined as the
fraction of connections between topological neighbors of any
site\cite{AlBa2002}. In Fig. \ref{fig3} we show the typical
behavior of $L(n)$ and $Cc(n)$. We see that  $Cc(n) \sim
n^{-0.75}$ and $L(n) \sim A\, \ln{n} + C$. Such scaling behavior
is the same one observed in the BA model\cite{AlBa2002}.

As shown in Fig.~\ref{fig3} larger networks becomes less clustered and have
longer minimum path on the average.
 $L$ and $Cc$ are inversely
related as it can be seen in the single run plotted in panel b of
Fig.~\ref{fig3}. The data correspond to values computed every fifty
trials, whether or not the attempt to add a node was successful
or not at that trial. In a sense, this is how a natural network
would look like to an observer if one could take snapshots in
time. Clearly both quantities fluctuate in opposite directions, as
further shown in the inset where the data corresponding to a
randomly shuffled network is also plotted for comparison. The
behavior of $Cc$ and $L$ is linked with the selection dynamics
ruling which node is accepted or rejected. The stability
constraint favors the nodes with few links, since they modify the
matrix ${a_{i,j}}$ stability much less than new nodes with many
links (of course this is reflected in the $P(k)$ density). Thus,
most frequently the network grows at the expense of adding nodes
with one or few links, producing an increase of $L$ and a
decreases of $Cc$. Most of the times, nodes with many links
destabilize the network and are rejected, but when one is finally
accepted, a large decrease in $L$ together with an increase in
$Cc$ is observed. This sudden change is the signature of a new
network hub, as seen in the example denoted with an arrow in Fig.
4b. We also verified that those fluctuations lead to a slow
diffusive-like growth of the network size $n(t)\sim t^{1/2}$ (not shown),
where the time is measured in number of trials.

\begin{figure}
\begin{center}
\includegraphics[scale=0.28]{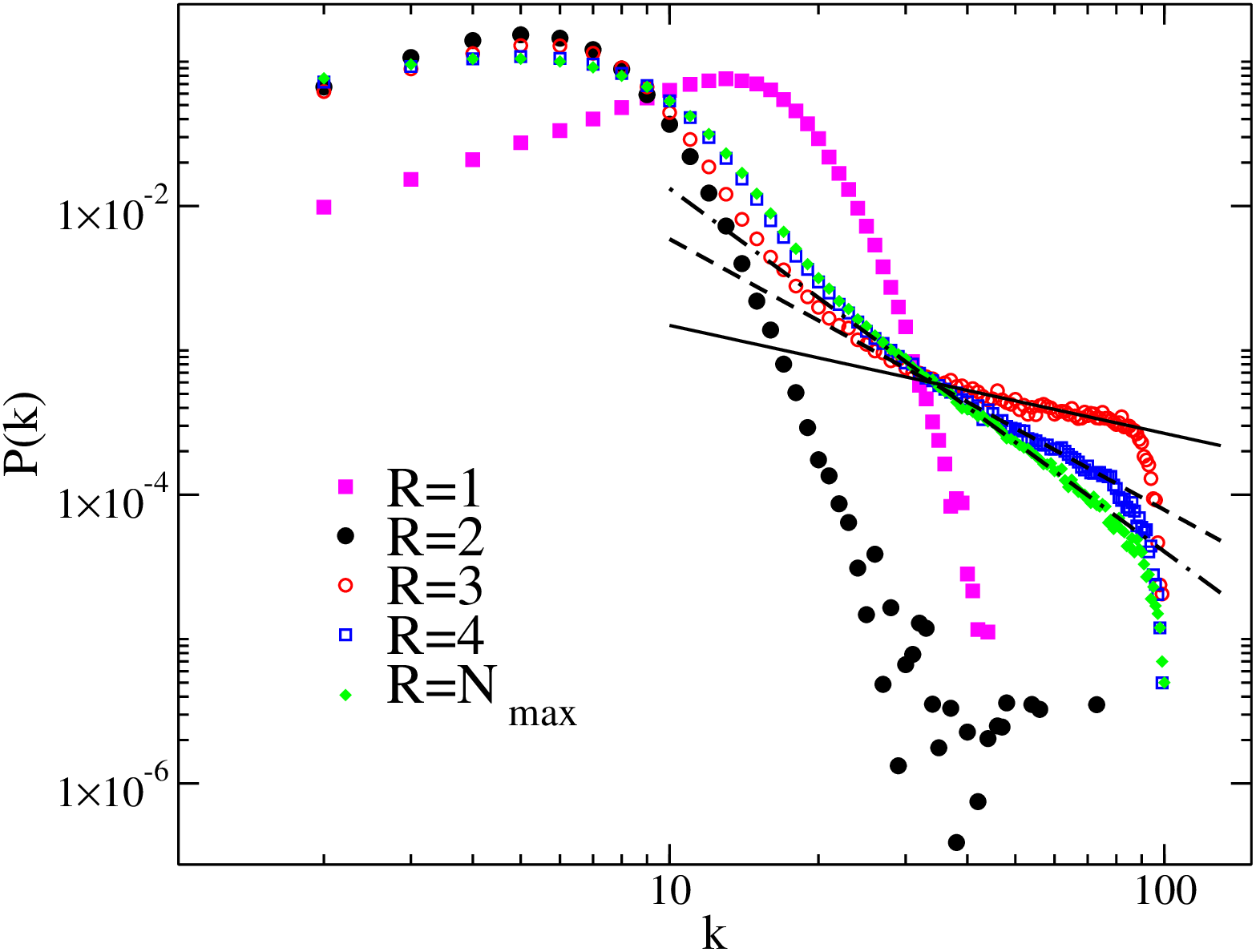}
\caption{\label{fig4} (Color on-line)  Degree distribution $P(k)$
for $b=2$, $q=3$, $n_{max}=100$ and  different values of $R$; the straight
lines correspond to a power law fits.}
\end{center}
\end{figure}

Finally, to consider the effect of {\it local} stability selection pressure, we modified the algorithm as follows. Once the new candidate node and its 1st nearest neighbors (nn) are chosen, we analyze the stability of the subnetwork composed by neighbors up to a range $R$ ($R=1$: 1st nn, $R=2$: 2nd nn, etc.). In Fig.\ref{fig4} we show $P(k)$ at $n=n_{max}$ for different values of $R$. We see that the fat tail $P(k)\sim k^{-\gamma}$ appears as long as $R\geq 3$, which coincides with the value of $L$ for the corresponding net size (see Fig. 3), suggesting a  correlation between stability and the self organized emergence of small world topology. Notice also that considering local stability allows a larger variability in the value of $\gamma$ ($\gamma\approx 0.9$ for $R=3$), although $\gamma$ quickly converges to the global stability result (for $R>4$ both results are almost indistinguishable).

The robustness of complex networks against error and attack has
already been investigated\cite{AlJeBa2000} considering the effects of nodes or
links' deletion. The present results shows that the consequences of
perturbing a single node may depend also on stability, a topic that
deserves further clarification.
Indeed, closely related results on  Boolean networks dynamics supports already the generality of this approach\cite{AlCl2003}.

Summarizing, the analysis of a  simple model shows that
complex topology can appear in networks as an
emergent property driven by a stability
selection pressure during the growing process.
This suggests yet another explanation for the
ubiquity of complex topology observed in different networks
in nature.

 Work
supported by CONICET, Universidad Nacional de
C\'ordoba, FONCyT grant PICT-2005 33305 (Argentina)
and NINDS (USA).


\begin{thebibliography}{9000}
\bibitem{NoteLett1} This rule is a bit restrictive, in
the  sense that one should rather expect an avalanche of
extinctions once the system destabilizes. However  this  minimal
network model is already enough to demonstrate scale free
topology.
\bibitem{numerics}The process is repeated for $n=n_{max}$ (typically $n_{max}=200$)
 and restarted $10^5$ times from $n=1$ to obtain statistics of the networks
 properties for different values of $b$ and $q$.
\end{thebibliography}
\end{document}